\documentclass[aps,prl,reprint,showpacs,longbibliography,superscriptaddress]{revtex4-1}
\usepackage{graphicx,color,bm}
\usepackage{epstopdf} % for PDFlatex

\newcommand{\RFM}{{RbFe(MoO$_4$)$_2$}}
\newcommand{\RKFM}{{Rb$_{1-x}$K$_{x}$Fe(MoO$_4$)$_2$}}
\newcommand\Hsat{$H_{\rm sat}$}

\newcommand{\be}{\begin{equation}}
\newcommand{\ee}{\end{equation}}
\newcommand{\bea}{\begin{eqnarray}}
\newcommand{\eea}{\end{eqnarray}}

\begin{document}
\title{Order by quenched disorder in the model triangular antiferromagnet RbFe(MoO$_4$)$_2$}

\author{A.~I.~Smirnov}  \affiliation{P.~L.~Kapitza Institute for Physical Problems, RAS, 119334 Moscow, Russia}
\author{T.~A.~Soldatov} \affiliation{P.~L.~Kapitza Institute for Physical Problems, RAS, 119334 Moscow, Russia}
                    \affiliation{{Moscow Institute for Physics and Technology, 141700 Dolgoprudny, Russia}}
\author{O.~A.~Petrenko} \affiliation{Department of Physics, University of Warwick, Coventry CV4 7AL, United Kingdom}
\author{A.~Takata}      \affiliation{Center for Advanced High Magnetic Field Science (AHMF), Osaka University, Osaka 560-0043, Japan}
\author{T.~Kida}        \affiliation{Center for Advanced High Magnetic Field Science (AHMF), Osaka University, Osaka 560-0043, Japan}
\author{M.~Hagiwara}    \affiliation{Center for Advanced High Magnetic Field Science (AHMF), Osaka University, Osaka 560-0043, Japan}
\author{A.~Ya.~Shapiro} \affiliation{A. V. Shubnikov Institute for Crystallography RAS, 119333 Moscow, Russia}
\author{M.~E.~Zhitomirsky}  \affiliation{CEA, INAC-PHELIQS, F-38000 Grenoble, France}

\date{\today}

\begin{abstract}
We observe a disappearance of the 1/3 magnetization plateau and a striking change of the magnetic configuration
under a moderate doping of the model triangular antiferromagnet \RFM. The reason is an effective lifting of
degeneracy of mean-field ground states by a random potential of impurities, which compensates, in the low
temperature limit, the fluctuation contribution to free energy. These results provide a direct experimental
confirmation of the fluctuation origin of the ground state in a real frustrated system. The change of the ground
state to a least collinear configuration reveals an effective positive biquadratic exchange provided by the
structural disorder. On heating, doped samples regain the structure of a pure compound thus allowing for an
investigation of the remarkable competition between thermal and structural disorder.
\end{abstract}

\maketitle

Triangular-lattice antiferromagnets (TLAF) form a popular class of magnetic materials that provides paradigmatic
cases of magnetic frustration~\cite{Petrenko} and intrinsic multiferroicity~\cite{Kenzelman07,Lewtas10}. A
spectacular manifestation of frustration in TLAFs is the 1/3 magnetization plateau with the up-up-down ($uud$)
spin structure stabilized by fluctuations out of a degenerate manifold of classical ground
states~\cite{Lee84,Kawamura,Chubukov91}. A 1/3 magnetization plateau has been experimentally observed in a number
of TLAFs~\cite{Inami96,SvistovPRB2003,Kitazawa99,Ishii11,ShirataPRL}. Generally, it is a challenging task to
distinguish the pure frustration-driven plateaus from those produced by additional perturbations, such as an
Ising anisotropy or magnetoelastic coupling. The latter mechanism is responsible for the 1/2 magnetization
plateau in chromium spinels~\cite{Ueda05,Penc04,Ueda06}, whereas an Ising anisotropy may contribute to the 1/3
plateau in some triangular antiferromagnets~\cite{Kitazawa99,Ishii11}. Thus, a direct verification of the plateau
mechanism remains a pressing issue in the field of frustrated magnetism.

An experimental test to determine if fluctuations are at the origin of the magnetization plateau states in TLAFs
has recently been suggested~\cite{Maryasin13}. It was shown theoretically that the frustration-driven plateaus
become unstable in the presence of a weak structural disorder either in the form of nonmagnetic dilution or as an
exchange-bond randomness. Weak disorder in frustrated magnets makes an energetic selection among degenerate
ground states that competes with the effect of thermal or quantum fluctuations. Besides the plateau smearing, in
fields below the plateau, the structural disorder may also stabilize umbrella- or fan-type magnetic structures
instead of the more collinear
states selected by fluctuations in clean TLAFs~
\cite{Lee84,Kawamura,Chubukov91}. The strong influence of a weak doping is due to a high degree of
degeneracy in a magnet with a fluctuation-selected ground state and will not give a comparable effect in systems
with a pronounced minimum of the mean-field energy. Thus the experimental observation of the plateau vanishing
and of a ground state changing on doping serves as direct evidence of the fluctuation origin of corresponding
phases in pure crystals with magnetic frustration. At the phenomenological level, the effect of structural disorder
on degeneracy lifting in frustrated magnets may be represented by an effective biquadratic exchange with a
positive sign~\cite{Henley89,Long89,Maryasin13,Fyodorov91,Maryasin15}. In a different context, a positive biquadratic exchange was shown to be
generated by surface roughness in ferromagnetic multilayers ~\cite{Slon91}, where it leads to the experimentally
observed 90$^\circ$ orientation of magnetizations ~\cite{Demokritov98,Schmidt1999}.

Our work is motivated by a search for direct experimental evidence for a disorder-induced positive biquadratic
exchange in bulk frustrated magnets. We also seek to verify experimentally the fluctuation nature of the ground
state of a pure TLAF and to observe the competition between static and dynamic disorder. The material chosen for
the study is \RFM, a magnetic compound featuring triangular-lattice layers of Fe$^{3+}$ ions with semiclassical
$S = 5/2$ spins. The system has an easy-plane magnetic anisotropy with the plane parallel to layers and exhibits
a 1/3 magnetization plateau only for $H\perp c$ (the $c$~axis is normal to the Fe$^{3+}$ layers)
~\cite{Svistov2006,Smirnov2007}. Random exchange-bond modulations in triangular layers is achieved by
substituting K for Rb in the interlayer space. In this Letter we present experimental results that demonstrate a
rapid disappearance of the plateau on K-doping, whereas the N\'{e}el temperature $T_{\rm N}$, the saturation
field \Hsat, and the antiferromagnetic resonance gap $\Delta$ are all changed by less than 15\%. Furthermore, the
electron spin resonance (ESR) measurements also reveal a drastic change of the spin structure with doping at
fields below the plateau.

%%%%%%%%%%%%%%%%%%%%%%%%%%%%%%%%%%%%%%%%%% Figure 1 %%%%%%%%%%%%%%%%%%%%%%%%%%%%%%%%%%%%%%%%%%%%%%%%%%
\begin{figure}[tb]
\centering
\includegraphics[width=\columnwidth]{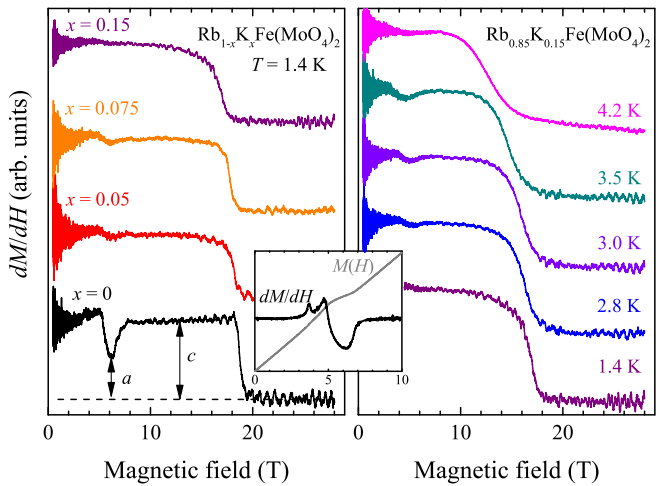}
\caption{(Color online) Left, $dM/dH$ {\it vs} field curves for the \RKFM\ crystals with different $x$, measured
in the pulse field  ${\bf H}\perp c$ at $T=1.4$~K.
    Parameters $a$ and $c$ are used to define a plateau {\it quality} factor $Q$ (see main text).
    Inset, $M(H)$ and $dM(H)/dH$ curves taken for a pure sample at the same temperature in steady fields using
    a vibrating sample magnetometer.
    Right, $dM/dH$ {\it vs} field curves for the $x=0.15$ sample measured in the pulsed field
    ${\bf H}\perp c$ at different temperatures.}
\label{Fig1_Magnet}
\end{figure}
%%%%%%%%%%%%%%%%%%%%%%%%%%%%%%%%%%%%%%%%%% Figure 1 %%%%%%%%%%%%%%%%%%%%%%%%%%%%%%%%%%%%%%%%%%%%%%%%%%

The crystals of \RKFM\ were prepared as previously described~\cite{SvistovPRB2003}. In addition to a controlled
amount of K, about 1 atomic percent of Al was also found in some samples. This aluminum impurity, probably comes
from the corundum crucible and is a likely reason for the observed dispersion in $T_{\rm N}$ of about $\pm 0.2$~K
for the samples with the same $x$. In the limiting case of  KFe(MoO$_4$)$_2$ ($x$=1) we get a $\sim 50$\%
reduction of the principal exchange constant~\cite{KFeMoOPRL}. The magnetization curves $M(H)$ and susceptibility
were studied by means of the vibrating sample magnetometer in magnetic fields up to 10~T and by a pulsed field
method in the 30~T range using the 55~T magnet at the AHMF center in Osaka University. Multifrequency 25--150~GHz
ESR was used to determine zero-field energy gap in AHMF center and to derive frequency-field diagrams and angular
dependencies of ESR absorption in Kapitza Institute.

The magnetization plateau for the pure sample (as illustrated in the inset of Fig.~\ref{Fig1_Magnet}) is clearly
marked by a significant reduction in the derivative, $dM/dH$, measured for ${\bf H}\perp c$. The development of
the $dM(H)/dH$ curves on doping is presented in the left panel of Fig.~\ref{Fig1_Magnet}. At $T=1.4$~K the
decrease in the $dM/dH$ value on the plateau becomes smaller with doping, and for $x=0.15$ the plateau completely
disappears. However, the temperature evolution of $dM(H)/dH$ curves for this $x=0.15$ sample shows, that the
plateau is restored, at least partially, on heating. The local minimum in the derivative appears again near 6~T
for $T=2.8$~K and remains clearly visible at 3.0 and 3.5~K, as shown in Fig.~\ref{Fig1_Magnet}, right panel. We
present here only the records of the $dM(H)/dH$ curves for decreasing magnetic field, as this sweep direction
helps to avoid the sample heating caused by a magnetocaloric effect. The full collection of the $dM(H)/dH$ curves
for both increasing and decreasing fields is given in Supplemental Material~[\onlinecite{Supplemental_material}].

%%%%%%%%%%%%%%%%%%%%%%%%%%%%%%%%%%%%%%%%%% Figure 2 %%%%%%%%%%%%%%%%%%%%%%%%%%%%%%%%%%%%%%%%%%%%%%%%%%
\begin{figure}[b]
\centering
\includegraphics[width=\columnwidth]{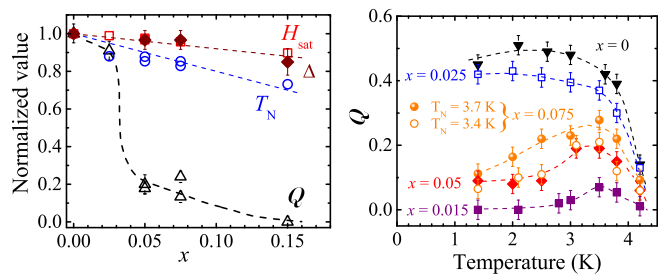}
\caption{(Color online) Left panel, normalized values of N\'{e}el temperature $T_{\rm N}$, saturation field
\Hsat, excitation gap $\Delta$ and plateau quality factor $Q$ {\it vs} doping.
    Right panel, plateau quality factor $Q$ {\it vs} temperature for  samples with different doping concentration.
    Dashed lines are guides to the eyes.}
\label{Fig2_Q_TN}
\end{figure}
%%%%%%%%%%%%%%%%%%%%%%%%%%%%%%%%%%%%%%%%%% Figure 2 %%%%%%%%%%%%%%%%%%%%%%%%%%%%%%%%%%%%%%%%%%%%%%%%%%

We introduce a {\it quality} factor, $Q=1-a/c$, to characterize the magnetization plateau. Here $a$ is the value
of the $dM/dH$ at its minimum on a plateau and $c$ is the value of the $dM/dH$ for the fields between the plateau
and magnetization saturation, as defined in Fig.~\ref{Fig1_Magnet}. For an ideal, perfectly flat plateau $Q=1$,
while $Q=0$ corresponds to the absence of the plateau. The dependence of $Q$ on doping and temperature (see
Fig.~\ref{Fig2_Q_TN}) shows a disappearance of the plateau on doping and its restoration on heating. The value of
$H_{\rm sat}$ is defined as a field where $dM/dH$ decreases by 50\% compared to its value just above a plateau.
The value of $\mu_0 H_{\rm sat}$ measured at $T=1.4$~K decreases on doping from 18.6~T in a pure sample to 16.7~T
for $x=0.15$. An empirical width of the field transition to a saturation, estimated as a field interval where the
$dM/dH$ varies between 0.75 and 0.25 of the maximum value is 0.3~T for the $x=0$ sample, and it increases to
about 1.0~T for the $x=0.15$ sample, but is still much smaller than the saturation field itself. Using a similar
criterion, that the change of derivative $dM/dT$ varies between 0.75 and 0.25 of the maximum value, the
transition width $\Delta T$ at $T_{\rm N}$ can also be estimated. $\Delta T$ is about 0.1~K for all $x$, except
$x=0.05$ where it is 0.3~K, thus showing that the transition region is significantly smaller than $T_{\rm N}$. On
doping, the value of $T_{\rm N}$ decreases from 4.1 for $x=0$ to 3.0~K for the $x=0.15$ sample. The
$x$-dependencies of the normalized values of $T_{\rm N}$ and \Hsat\ are presented in the left panel of
Fig.~\ref{Fig2_Q_TN}.

%%%%%%%%%%%%%%%%%%%%%%%%%%%%%%%%%%%%%%%%%% Figure 3 %%%%%%%%%%%%%%%%%%%%%%%%%%%%%%%%%%%%%%%%%%%%%%%%%%
\begin{figure}[tb]
\centering
\includegraphics[width=\columnwidth]{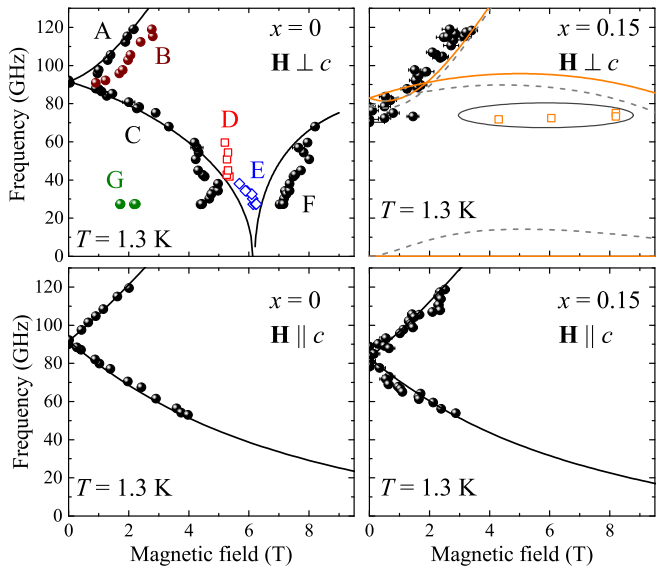}
\caption{(Color online) Frequency-field diagrams of the \RKFM\ samples for $x=0$ (left-hand side panels) and
$x=0.15$ (right-hand side panels).
    The top and bottom panels show the ESR frequencies for $H\perp c$ and $H\parallel c$, respectively.
    The experimental results (symbols) are compared to the calculations (solid lines) for the Y/umbrella structure with $J=1.1J_0$, $D=D_0$ for $x=0$ and for the inverted Y ($\overline{\rm Y}$)/umbrella state with $J=0.9J_0$, $D=D_0$ for $x=0.15$, see main text for the notations.
    Dashed lines represent a calculation for the $\overline{\rm Y}$ state with $K/J=0.01$, while the ellipse covers a wide area of a weak absorption.}
    \label{Fig3_ESR}
\end{figure}
%%%%%%%%%%%%%%%%%%%%%%%%%%%%%%%%%%%%%%%%%% Figure 3 %%%%%%%%%%%%%%%%%%%%%%%%%%%%%%%%%%%%%%%%%%%%%%%%%%

The ESR spectra of a nominally pure sample (left panels of Fig.~\ref{Fig3_ESR}) have an energy gap of $\Delta_0
\simeq 90$~GHz and for $H \perp c$ consist of the two frequency branches, ascending and descending with an
applied magnetic field in agreement with the previous results~\cite{SvistovPRB2003,Smirnov2007}. The ascending
branch is split into two closely-positioned branches, (A) and (B), by the weak interplane interactions. The
frequency of the descending branch, (C), is reduced almost to zero for the field approaching the lower boundary
of the plateau, while another mode, (D), of a higher frequency, appears near the entrance of the plateau. Modes
in the middle of the plateau (E) and at the upper boundary (F) are also detected in qualitative agreement with
the theory~\cite{Chubukov91} and previous experiments~\cite{SvistovPRB2003}. Finally, a weakly-absorbing mode (G)
near 30~GHz appears due to a splitting of the zero-frequency mode~\cite{SvistovPRB2003}. A full record of the
microwave absorption {\it vs} magnetic field at different frequencies is given in Supplemental
Material~\cite{Supplemental_material}.

The frequency-field dependencies for the $x=0.15$ sample are shown in the right-hand panels of
Fig.~\ref{Fig3_ESR}. The gap $\Delta$ is reduced to $(75 \pm 5)$~GHz on doping, its evolution with doping
concentration $x$ is shown in Fig.~\ref{Fig2_Q_TN}. For a pure sample, ESR lines are observed both above and
below $\Delta_0$ for ${\bf H} \perp c$ while for the $x=0.15$ sample only the ascending ESR branch is detected.
The descending ESR branch either disappears or transforms in a field-independent mode on doping. For the field,
${\bf H} \parallel c$, both the ascending and descending branches are visible for pure and all doped samples, see
lower panels in Fig.~\ref{Fig3_ESR} and~\cite{Supplemental_material}. The angular dependence of the ESR
absorption, presented in~\cite{Supplemental_material} reveals, that upon rotating from ${\bf H} \parallel c$  to
${\bf H} \perp c$, the ESR line of the descending branch is smeared at a deviation from the $c$~axis and
disappears completely for ${\bf H} \perp c$ in the $x=0.15$ sample, while it is conserved in a pure sample. This
observation gives a direct confirmation of the disappearance of the descending ESR branch on doping.

%%%%%%%%%%%%%%%%%%%%%%%%%%%%%%%%%%%%%%%%%% Figure 4 %%%%%%%%%%%%%%%%%%%%%%%%%%%%%%%%%%%%%%%%%%%%%%%%
\begin{figure}[b]
\centerline{
\includegraphics[width=0.7\columnwidth]{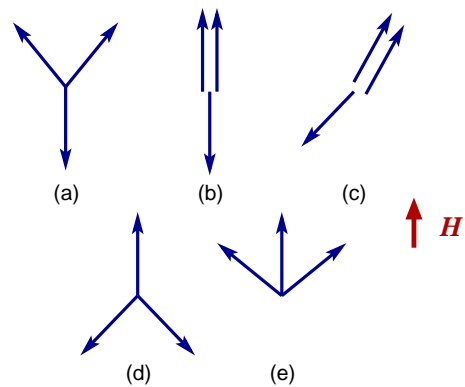}}
\caption{Spin structures appearing for the pure TLAF in an applied field: (a) the Y state, (b) the $uud$ state,
(c) the 2:1 state.
    Spin structures of a planar triangular antiferromagnet in the presence of structural disorder: (d) $\overline{\rm Y}$ state and (e) the fan state.}
\label{Fig4_spinstructure}
\end{figure}
%%%%%%%%%%%%%%%%%%%%%%%%%%%%%%%%%%%%%%%%%% Figure 4 %%%%%%%%%%%%%%%%%%%%%%%%%%%%%%%%%%%%%%%%%%%%%%%%

To model the observed behavior of doped \RFM, we use the spin Hamiltonian
\begin{equation}
\hat{\cal H} = \sum_{\langle ij\rangle} J_{ij} {\bf S}_i\cdot {\bf S}_j + D \sum_i (S_i^z)^2 - g\mu_B\sum_i{\bf
H}\cdot{\bf S}_i \ . \label{eq:H}
\end{equation}
The nearest-neighbor exchange  constant $J_{ij}$ is assumed to have weak random variation between the bonds:
$J_{ij} = \langle J_{ij} \rangle + \delta J_{ij}$, $\langle \delta J_{ij} \delta J_{kl} \rangle = \delta J^2
\delta_{ij,kl}$, $\delta J \ll J =  \langle J_{ij} \rangle$. Inelastic neutron scattering experiments yield
$J_0=0.086$~meV and $D_0=0.027$~meV for \RFM~\cite{White13}. The effect of bond disorder and/or thermal and
quantum fluctuations can be semi-quantitatively represented by adding a biquadratic term
\begin{equation}
\hat{\cal H}_B = K \sum_{\langle ij\rangle} \bigl({\bf S}_i\cdot {\bf S}_j\bigr)^2 \ . \label{eq:BQ}
\end{equation}
For the purely Heisenberg model (\ref{eq:H}) with $D=0$, $\delta J_{ij}=0$, the quantum fluctuations contribute
$K_Q \simeq - J/(24S^3)$~\cite{MZH15}. A negative sign for the biquadratic constant $K$ means that fluctuations
select the most colinear/coplanar states resulting in a standard sequence of the ordered states as the field
increases, see Fig.~\ref{Fig4_spinstructure}(a)--(c)~\cite{Chubukov91}. A frozen bond disorder instead generates
a positive biquadratic contribution, $K_{\rm dis} = \delta J^2/(3JS^2)$~\cite{Maryasin13}. For $(\delta J/J)^2>
1/(8S)$, the disorder effect becomes dominant and the stable magnetic states are found by minimizing a
Hamiltonian, which combines (\ref{eq:H}) and (\ref{eq:BQ}) with $K>0$. One can straightforwardly verify that in
such a case the least collinear states are energetically preferred. For the planar spin system ($D>0$), this is
the inverted Y ($\overline{\rm Y}$) state, Fig.~\ref{Fig4_spinstructure}(d), which continuously transforms into
the fan state in higher fields, Fig.~\ref{Fig4_spinstructure}(e).
Small variations in the direction of the local anisotropy axis may also contribute to a lifting of the degeneracy and $K_{dis}$, but this weak effect is ignored as $D<J$.

The two possible low-field magnetic structures, the Y and the $\overline{\rm Y}$ states, have qualitatively
different ESR spectra for in-plane orientations of an applied field. This fact fully agrees with the
idea of an order-by-disorder effect, which relates the lifting of degeneracy in frustrated magnets to the
excitation spectra of the different ground states~\cite{Shender82,Moessner01}. Performing the standard spin-wave calculations for
 %the Hamiltonian~
 (\ref{eq:H}) we obtain  two two resonance frequencies:
\begin{eqnarray}
\omega_1 & = & 3JS \sqrt{d(1\mp h)(3\pm h)} \ , \ \ d = \frac{D}{3J}     \nonumber \\
\omega_2 & = & 3JS \sqrt{2d + h^2 + d(1\pm h)^2}\ , \ \ h = \frac{g\mu_B H}{3JS}  \ , \label{W12}
\end{eqnarray}
where the upper/lower sign corresponds to the Y/$\overline{\rm Y}$ state. The third ESR branch vanishes in the
harmonic approximation $\omega_3=0$ reflecting the remaining classical degeneracy.

For $H\perp c$, the ESR frequencies of pure \RFM\ shown in the upper left panel of Fig.~\ref{Fig3_ESR}, are
accurately described by the expressions (\ref{W12}) for the Y state with $J = 1.1J_0$ and $D = D_0$, the values
that are marginally different from the ones obtained in the neutron scattering experiments~\cite{White13}. The
agreement between theory and experiment is good despite presence of the incommensurate spiral state in low fields
\cite{Kenzelman07}, which may indicate significant Y-type distortions of the spiral structure induced by an
external field. The same microscopic parameters also nicely fit the ESR data for $H\parallel c$, the lower left
panel of Fig.~\ref{Fig3_ESR}, see also~\cite{Supplemental_material}. The characteristic feature of the ESR
spectra for the Y state ($H\perp c$) is the descending branch $\omega_1(H)$ that corresponds to the out-of plane
oscillations of the two spin sublattices about the field direction, while the third sublattice remains parallel
to the field. The frequency of this mode decreases to zero upon a transition into the collinear $uud$
structure at $\mu_0 H\approx 6$~T. In contrast, the frequency of the same mode for the $\overline{\rm Y}$ state,
which remains noncollinear, exhibits little variation in the corresponding field region. Thus, the absence of the
descending ESR branch for the doped samples for $H\perp c$ clearly indicates a change of the spin structure. We
compare the ESR data for the $x=0.15$ doped sample with the theoretical frequencies (\ref{W12}) for the
$\overline{\rm Y}$ state on the upper right panel in Fig.~\ref{Fig3_ESR}. A somewhat smaller averaged value of $J
= 0.9J_0$ ($D = D_0$) used for the fit is consistent with a local reduction of the exchange interaction induced
by K impurities.

The biquadratic interaction (\ref{eq:BQ}) has been derived as an effective potential in the manifold of
degenerate classical ground states~\cite{MZH15}. Nevertheless, the effect of structural disorder on the $q=0$
excitations in the $\overline{\rm Y}$ state beyond substituting averaged $J$ and $D$ can be estimated by
explicitly including  $\hat{\cal H}_B$ in the calculations. One should bear in mind that an effective biquadratic
interaction is rather weak. Using $K_{\rm dis} = \delta J^2/(3JS^2)$~\cite{Maryasin13} and noting that the local
$\delta J/J \simeq 50$\%, we find that the biquadratic constant does not exceed $K_{\rm dis}/J=0.01$--0.02,
whereas $K_Q/J\simeq -3 \times 10^{-3}$. We show in the upper right panel of Fig.~\ref{Fig3_ESR} the ESR modes
computed for $J = 0.9J_0$, $D = D_0$, and $K/J=0.01$. The main qualitative effect of $\hat{\cal H}_B$ is a
nonzero value for $\omega_3$, which reflects a lack of degeneracy, whereas the shift of the two upper modes is
indeed tiny. The resonance frequencies for the third branch are too small to be detected in the ESR spectrometers
used here.

One can see in Fig.~\ref{Fig2_Q_TN}, that the plateau quality $Q$ changes drastically with doping, whereas
$T_{\rm N}$, \Hsat\ and $\Delta$ exhibit a weak linear dependence on $x$. The samples with the plateau suppressed
or canceled by the impurities still demonstrate a sharp N\'{e}el transition, shifted down in temperature from
$T_{\rm N}$ in a pure sample. The behavior of $T_{\rm N}$, \Hsat\ and $\Delta$ shows that the exchange
interaction is not strongly affected on doping. A sharp N\'{e}el transition and a step-like change in $dM/dH$ at
saturation confirm the absence of the macroscopic inhomogeneities in the samples studied. Thus, the observed
disappearance of the plateau and the change of the ground state in low fields should  only be ascribed to the
influence of a random potential. This statement is confirmed by the observed partial restoration of a plateau in
doped samples on heating (see Fig.~\ref{Fig2_Q_TN}). Indeed, according to Ref.~[\onlinecite{Maryasin13}], the
region of the $uud$-phase (1/3-plateau phase) in the $T-H$ phase diagram is shifted to a higher temperatures on
doping.

In conclusion, the observed doping-induced changes of the magnetization curves and the magnetic resonance spectra
of \RFM\ reveal a transition from a collinear up-up-down structure, stabilized by fluctuations, to a fan
structure supported by a weak static disorder, as well a transformation of the lower-field spin structure from
the Y-type to an inverted Y-structure. Our experiments establish the fluctuation origin of the 1/3-plateau and
the Y-type phases and show that the ground state selection process is affected by a strong competition between
structural disorder and thermal fluctuations. The structural disorder is found to lead to a positive biquadratic
exchange. We observe a fundamentally different behavior between pure and lightly doped samples on heating, which
results in the restoration of the magnetization plateau in the doped materials, while in a pure crystal the
plateau is removed. These observations provide convincing confirmation of the competition between thermal and
quenched disorder, demonstrating that the negative biquadratic term arising from thermal fluctuations once again
dominates at higher temperature. Disorder-induced modifications of the magnetic structure may also be used to
control multiferroicity of TLAFs and, perhaps, of other spiral antiferromagnets.

We thank S.S.~Sosin and L.E.~Svistov for numerous discussions. Work at the Kapitza Institute is supported by
Russian Foundation for Basic Research, grant No. 15-02-05918, by the Programs of the Presidium of Russian
Academy of Sciences, high-frequency ESR measurements are supported by Russian Science Foundation grant No. 17-12-01505. AIS is indebted to Osaka University for invitation as a visiting Professor. Work at Osaka University is supported by  the International Joint Research Promotion Program.

\onecolumngrid
\newpage
\setcounter{figure}{0}
\setcounter{equation}{0}

\twocolumngrid

{\bf{Supplemental Material for \\
 ``Order by quenched disorder in the triangular antiferromagnet  RbFe(MoO$_4$)$_2$''}}

\section{Sample preparation and characterization} \label{samples}
%%%%%%%%%%%%%%%%%
The crystals of \RKFM\ were prepared as previously described~[\onlinecite{PRB2003Svistov}].
The K content was determined  by means of energy-dispersive X-ray spectroscopy.
The crystal structure was checked by powder and single crystal X-ray diffraction and confirmed the crystal group as $P3\bar{m}1$ with the room-temperature lattice parameters $a=5.69$~\AA\ and $c=7.48$~\AA\ for $x=0$.
The $c$~axis exhibits a monotonic decrease with doping, for $x=0.15$ this decrease is $(0.030 \pm 0.015)$~\AA.

A small amount of aluminum impurity, about 1 atomic percent, was also found in some samples.
The aluminum impurity probably comes from the corundum crucible.
Being random and uncontrolled, it is the likely reason for the observed dispersion of about 0.2~K in the N\'{e}el temperature for the samples with nominally the same content of K.
For example, two nominally pure samples have different N\'{e}el temperatures: sample 1 with $T_{\rm N}=(4.1 \pm 0.05)$~K and sample 2 with $T_{\rm N}=(3.7\pm 0.05)$~K.
Similarly, for the two samples with $x=0.075$, the ordering temperature varies between 3.4 and 3.7~K.

\section{Experimental results}\label{Results}
\subsection{Magnetization curves}\label{dMdH}
%%%%%%%%%%%%%%%%%%%%%%%%
\begin{figure}[tb]
\centering
\includegraphics[width=0.85\columnwidth]{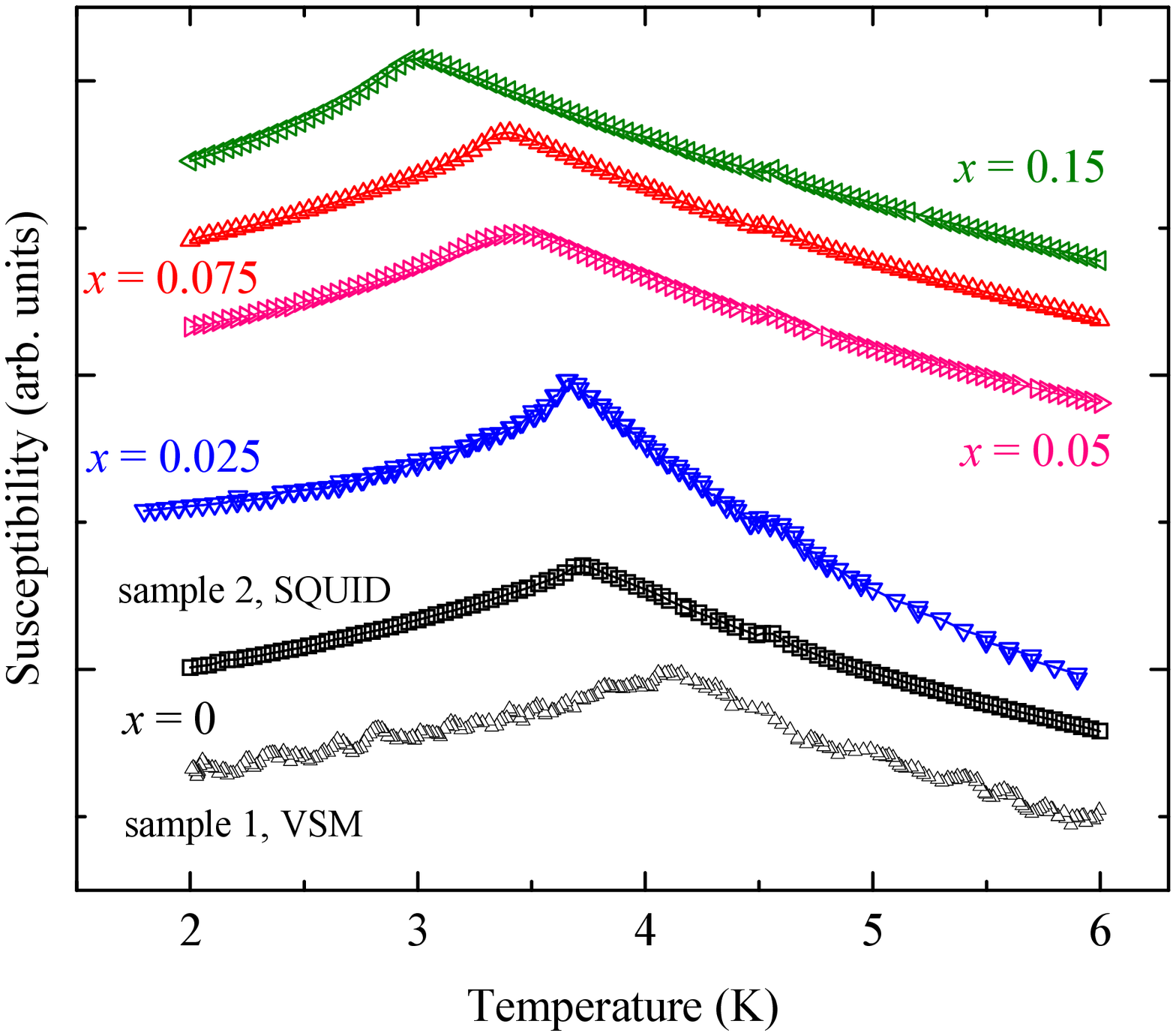}%%% FIG 1S %%%
 \caption{ Magnetic susceptibility {\it vs} temperature of \RKFM\  samples.
 		The data for the $x=0$, sample 1 are taken in a magnetic field of 0.05~T, while other curves are taken in a field of 0.1~T.
		The magnetic field is applied perpendicular to the $c$~axis.
		The curves are vertically offset for clarity.}
\label{fig:chivsT.eps}
\end{figure}
\begin{figure}[t]
\centering
\includegraphics[width=0.85\columnwidth]{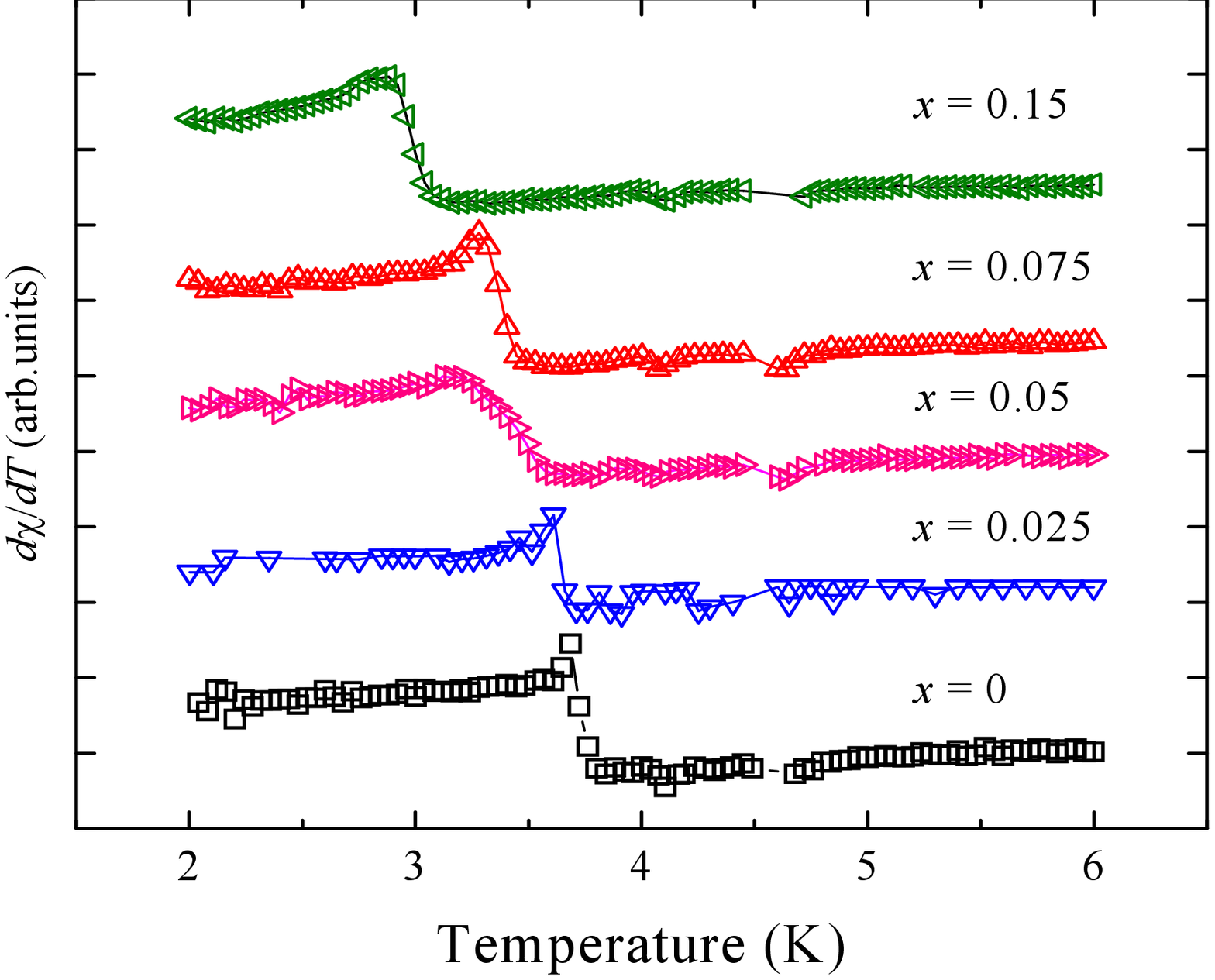}%%% FIG 2S %%%
 \caption{ Temperature derivative of the magnetic susceptibility, $d\chi/dT$ for the \RKFM\ samples with different doping concentration.
 		The curves are offset for clarity.}
\label{fig:dchidT}
\end{figure}
The susceptibility {\it vs} temperature, $\chi(T)$,  dependencies are shown in Fig.~\ref{fig:chivsT.eps}, the temperature dependence of the derivative $d\chi/dT$ is presented in Fig.~\ref{fig:dchidT}.
The susceptibility measurements were made with the magnetic field applied perpendicular to the $c$~axis, where a kink of  $\chi(T)$ clearly marks the N\'{e}el temperature $T_{\rm N}$.
The derivative $d\chi/dT$ has a negative sign at temperatures above the kink and is positive below the N\'{e}el point and, thus, shows a jump with a value of $\delta$ at the critical point.
It is clearly seen in Fig.~\ref{fig:chivsT.eps} that the N\'{e}el transition in doped samples is almost as sharp as in the pure samples.
We  estimate the width $\Delta T$ of the N\'{e}el transition as the temperature interval where the change of $d\chi/dT$  varies between 0.25 and 0.75 of the total jump $\delta$.
The width of the N\'{e}el transition is less than 0.3~K (see Fig.~\ref{fig:dchidT}) for all doped and pure samples, which is small in comparison to the N\'{e}el temperature itself.

\begin{figure}[b]
\centering
\includegraphics[width=0.95\columnwidth]{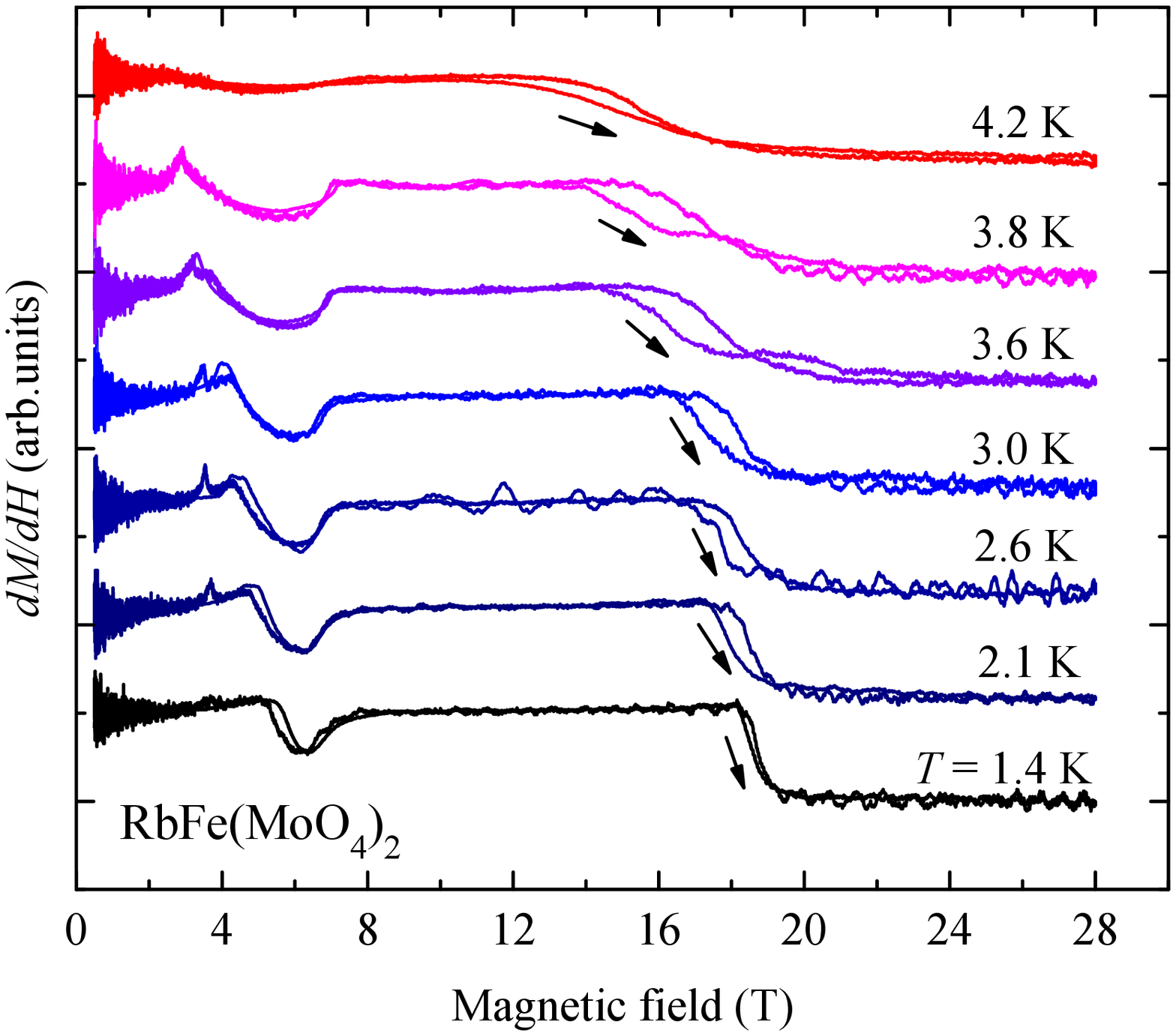}%%% FIG 3S %%%
 \caption{Magnetization derivative curves $dM/dH$ for a pure \RFM\ sample measured in a pulsed field for ${\bf H} \perp c$.
 		Both increasing and decreasing field records are shown.}
 \label{Fig:dMdH_pure}
\end{figure}

The 1/3 magnetization plateau, which is the remarkable feature of \RFM, is observable at the in-plane (${\bf H} \perp c $) orientation of the magnetic field, and is clearly marked by the drop of the derivative $dM/dH$ near the field of $H_{\rm sat}/3$, i.e. near 6~T.
The $dM/dH$ curves of a nominally pure sample with $T_{\rm N}=(4.1\pm0.05)$~K  recorded in a pulsed field in the whole range including the saturation field 18.6~T are presented in Fig.~\ref{Fig:dMdH_pure} and are analogous to those previously observed in Ref.~[\onlinecite{PRB2007Smirnov}].
The duration of the pulse of the magnetic field is 7~ms.

\begin{figure}[b]
\centering
\includegraphics[width=0.95\columnwidth]{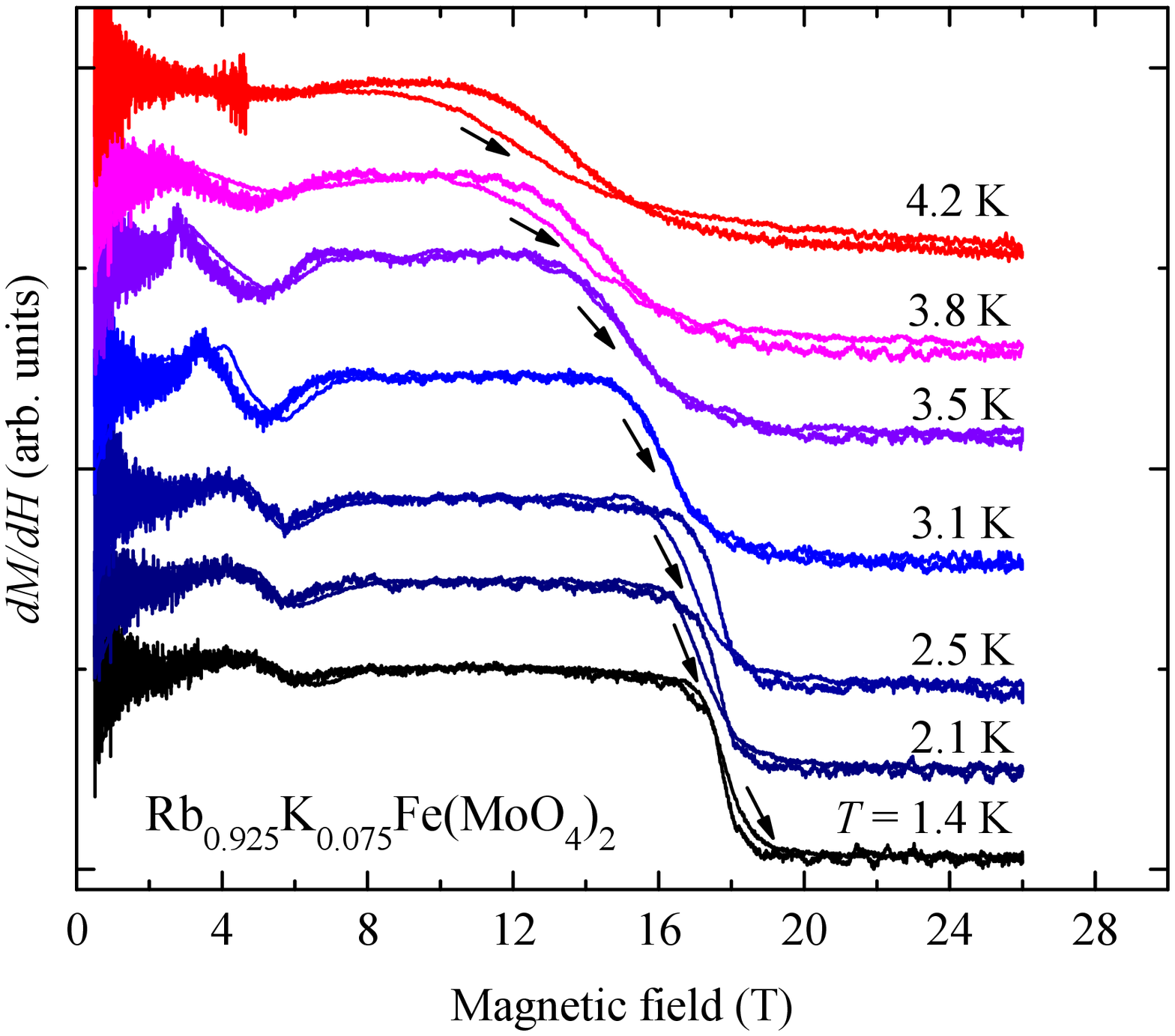}%%% FIG 4S %%%
 \caption{Magnetization derivative curves $dM/dH$ for the \RKFM\ sample with $x=0.075$ measured at different temperatures in a pulsed field for ${\bf H} \perp c$.
 		Both increasing and decreasing field records are shown.}
\label{Fig:dMdH_C0p075}
\end{figure}

The changes in the magnetization curves, which arise at intermediate doping of $x=0.075$, are shown in Fig.~\ref{Fig:dMdH_C0p075}: here we see, that at the lowest temperature 1.4~K, the drop of $dM/dH$ in the plateau range is much smaller than in the pure sample.
Nevertheless, the plateau still remains observable, in contrast to the sample with $x=0.15$, where it disappears completely at $T=1.4$~K (see main text).
The temperature dependencies of $dM/dH$ curves for $x=0.075$, presented in Fig.~\ref{Fig:dMdH_C0p075} demonstrate that the plateau, marked by a drop of the derivative near 6~T, is restored by an increase in the temperature.
At a temperature of about 3~K the plateau is again well pronounced: the drop of $dM/dH$ is about a half of that of a pure sample, while at $T=1.4$~K this drop is only one sixth of the drop in a pure sample.

The plateau recorded in the pulsed field  usually has a $Q$-value (see definition of $Q$ in the main text), which is 10-20\% lower than that recorded by the vibrating sample magnetometer in a steady field.
This is probably due to a finite time resolution of the recording system and a finite relaxation time of the spin system.
Sweeping through the field interval of  0.2~T at the plateau entrance, where $dM/dH$ is quickly changing, takes about 15 $\mu$s, this may be comparable or several times longer, than the relaxation time of the spin system of an antiferromagnet.
The hysteresis of the $dM/dH$ curves near the saturation field is due to the magnetocaloric effect, as described in Ref.~[\onlinecite{PRB2007Smirnov}].
Due to the magnetocaloric effect the sample is slightly heated during the magnetization process, while during the demagnetization it is cooled.
Because of this, we observe a larger saturation field when the field is reduced from the maximum value to zero, than during the field sweep from zero to the maximum value.
The saturation fields and values of the plateau quality $Q$ are quoted for the ``down"-records of $dM/dH$.

\subsection{Electron spin resonance}
\label{ESR}
\begin{figure}[tb]
\centering
\includegraphics[width=0.85\columnwidth]{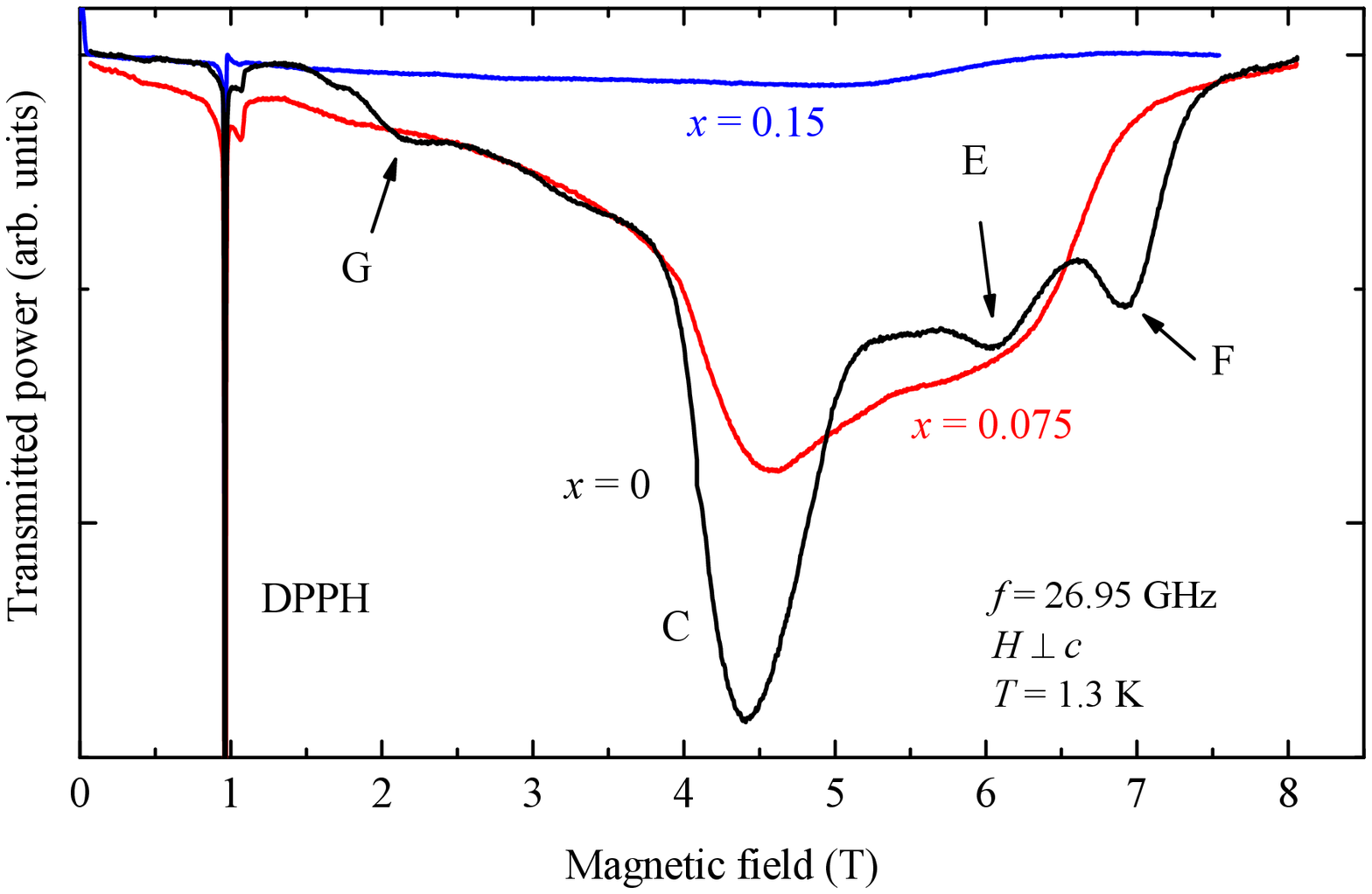}%%% FIG 5S %%%
\caption{Normalized ESR absorption curves (${\bf H} \perp c $) for the \RKFM\ samples with $x=0$, 0.075 and 0.15.}
\label{fig:27GHZESRinplane}
\end{figure}

%%%%%%%%%%%%%%%%%%%
Electron spin resonance (ESR) absorption lines were recorded as magnetic field dependencies of the microwave power, transmitted through the resonator with the sample.
The diminishing of this transmission indicates an increase in the absorption.
A small amount of  diphenyl-picryl-hydrazyl (DPPH) was used to mark the ESR field of free spins at the Larmor frequency $f_L=2\mu_B H/h$.
Figure \ref{fig:27GHZESRinplane} presents a comparison of the 26.9~GHz ESR absorption lines for a pure and two doped samples with $x=0.075$ and 0.15 at the lowest temperature $T=1.3$~K.
At this frequency the pure sample exhibits three resonances marked as C, E, F (the labels are the same as in the frequency-field diagram of Fig.~3 in the main text) in the plateau range, while for $x=0.075$ these modes are of much lower intensity and for $x=0.15$ they disappear completely, analogous to the plateau itself.

\begin{figure}[b]
\centering
\includegraphics[width=0.85\columnwidth]{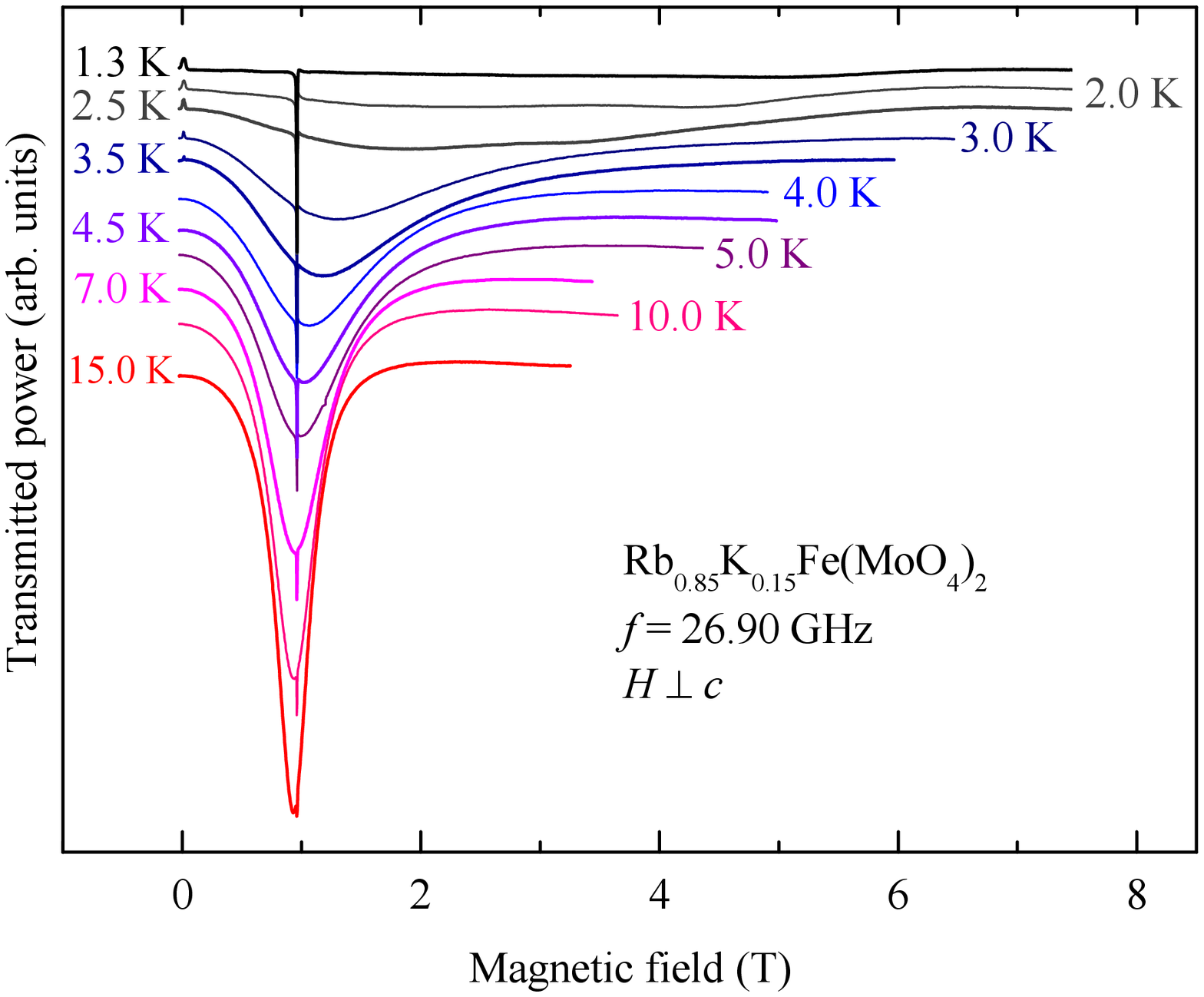}%%% FIG 6S %%%
 \caption{Temperature evolution of the ESR absorption curve for the \RKFM\ sample with $x=0.15$.
 		Narrow ESR signal in a magnetic field of 0.96~T is a DPPH marker.}
\label{fig:Tevol}
\end{figure}

Figure~\ref{fig:Tevol} presents the temperature evolution of the ESR absorption of the $x=0.15$ sample at a frequency of 26.9~GHz.
We see that the absorption features, suppressed by doping  for low temperature, are partially restored by heating above 2~K, analogous to the restoring of the plateau observed in the magnetization measurements described in the main text.
On heating above the N\'{e}el temperature, the ESR line transforms into a regular Lorentzian absorption curve at the resonance field of the Larmor precession 0.96~T.

\begin{figure}[tb]
\centering
\includegraphics[width=0.95\columnwidth]{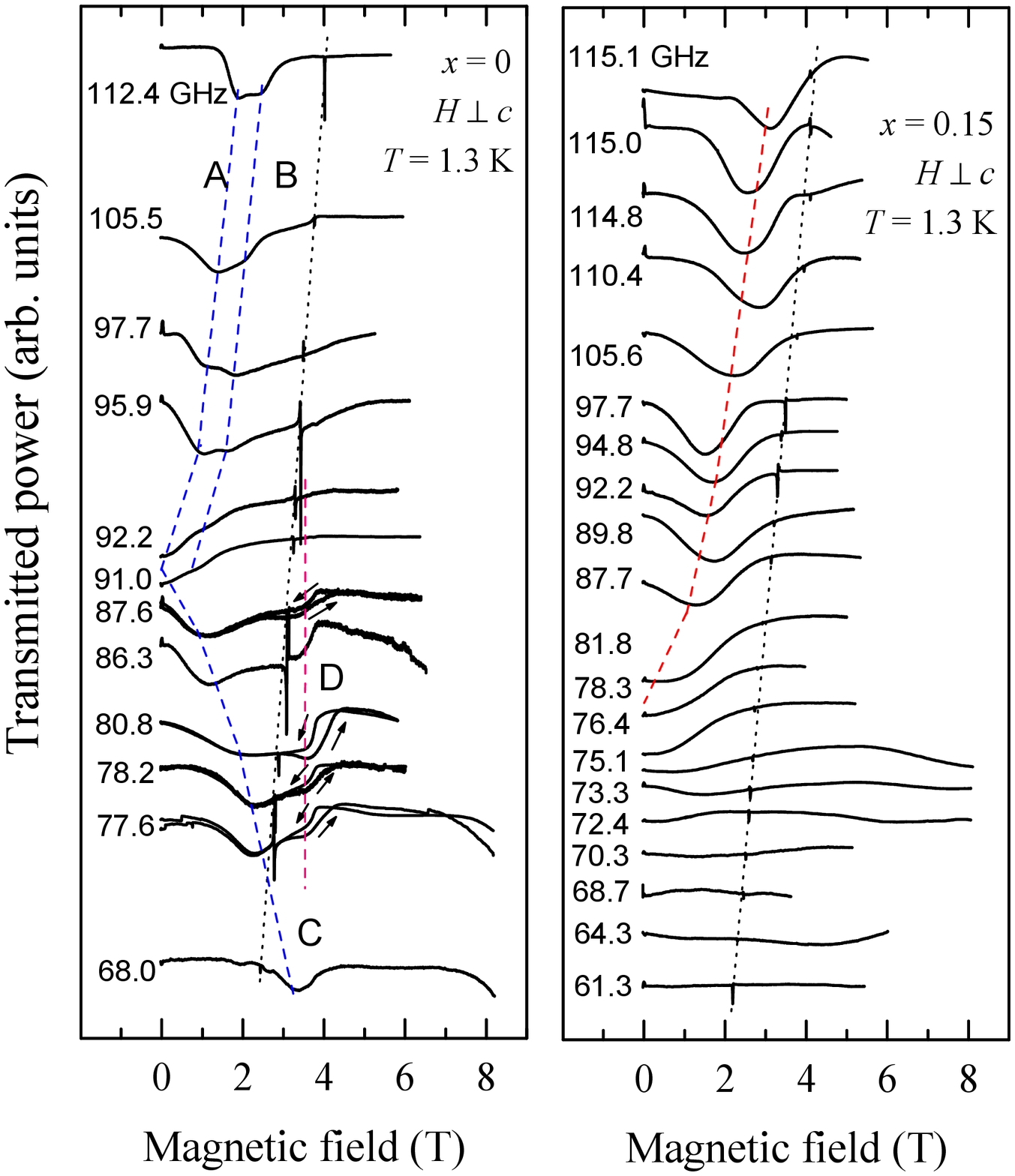}%%% FIG 7S %%%
\caption{Normalized ESR absorption curves for the in-plane magnetic field for \RKFM\ samples with $x=0$ and 0.15 for different frequencies.
		The gain is normalized to the intensity of the paramagnetic resonance signal at $T=10$~K.
		Dashed lines in left panel are guides to the eye: blue lines are for the AB-doublet of the ascending mode and for the descending mode C, black line is a DPPH label, red line is mode D at the right boundary of the plateau range.
		Dashed lines in the right panel are guides to the eye for the ascending branch (red line) and for a DPPH mark (black line).}
\label{fig:ESRlinesHinplanex0&x0p15}
\end{figure}

\begin{figure}[tb]
\centering
\includegraphics[width=0.85\columnwidth]{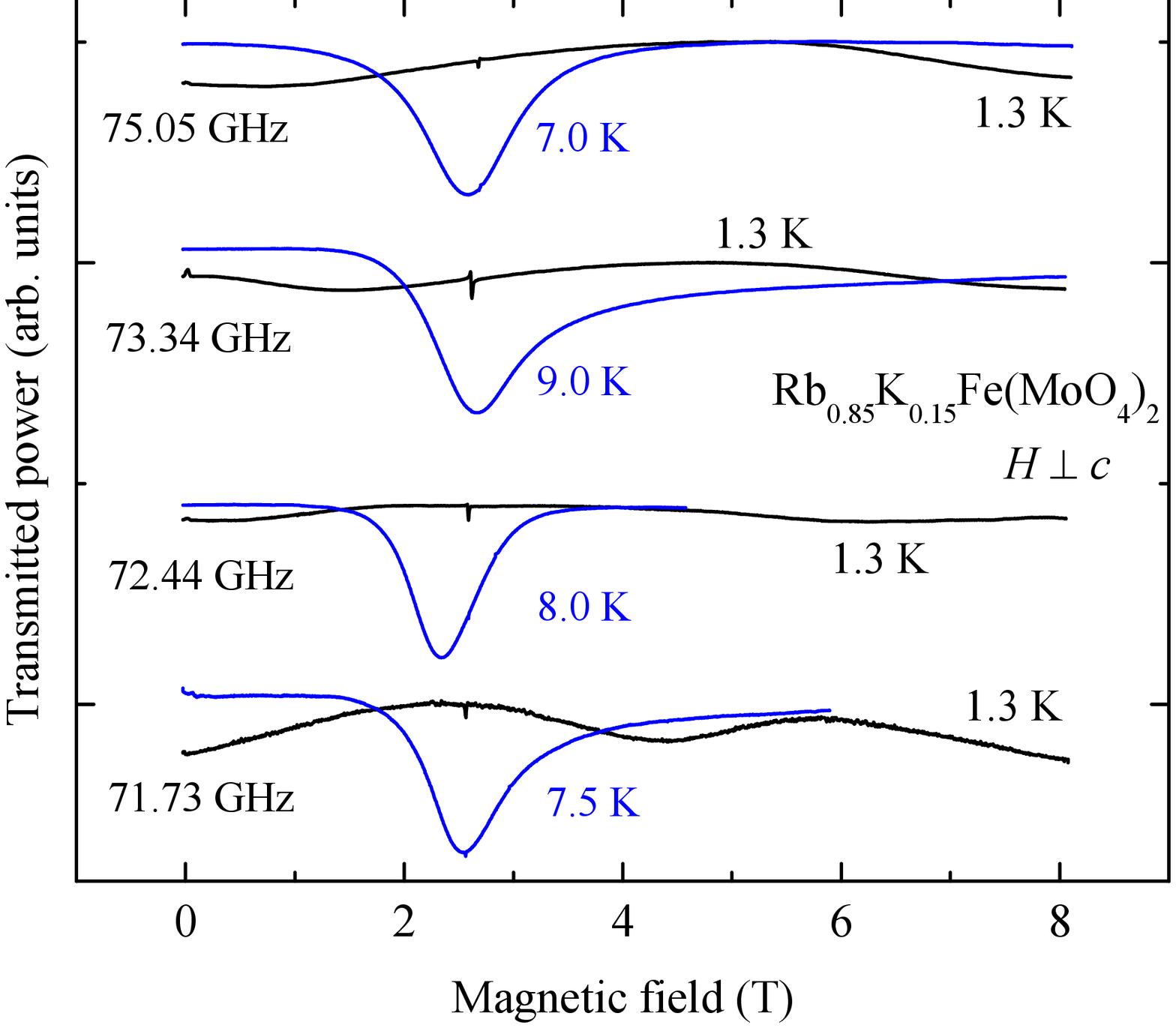}%%% FIG 8S %%%
 \caption{Weak microwave absorption in the vicinity of 74~GHz in the $x=0.15$ sample for ${\bf H} \perp c$ and $T=1.3$~K (black lines).
 		Blue lines show the  microwave transmission at higher temperature, with the same gain of a detector circuit.
		Narrow ESR signals are from a DPPH marker.}
\label{fig:wideabsorption}
\end{figure}

A comparison of ESR responses of pure and doped ($x=0.15$) samples at different frequencies $f$ is presented for the in-plane magnetic field in Fig.~\ref{fig:ESRlinesHinplanex0&x0p15}.
These records enable one to reconstruct the $f(H)$ dependencies, presented in Fig.~3 in the main text.
Figure~\ref{fig:ESRlinesHinplanex0&x0p15} illustrates the disappearance of the descending branch of the antiferromagnetic resonance spectrum in the $x=0.15$ sample.
Indeed, we can see that the ESR lines are clearly pronounced at frequencies below the 90~GHz gap for a pure sample.
At the same time, there are no visible resonance lines at the frequency below the 75~GHz gap for the $x=0.15$ sample.
On heating both the pure and the doped samples to a temperature of about 10~K, the Larmor frequency line of the same amplitude appears in both cases.
At a frequency  below the gap for the doped sample (75~GHz), the record of the microwave power transmitted through the resonator does not have a resonance shape and demonstrates only a weak variation of transmitted signal with magnetic field, see Fig.~\ref{fig:wideabsorption}.
This is to be compared with  pronounced ESR lines of the descending branch of the pure sample or of the doped sample for ${\bf H} \parallel c$.

\begin{figure}[tb]
\centering
\includegraphics[width=0.95\columnwidth]{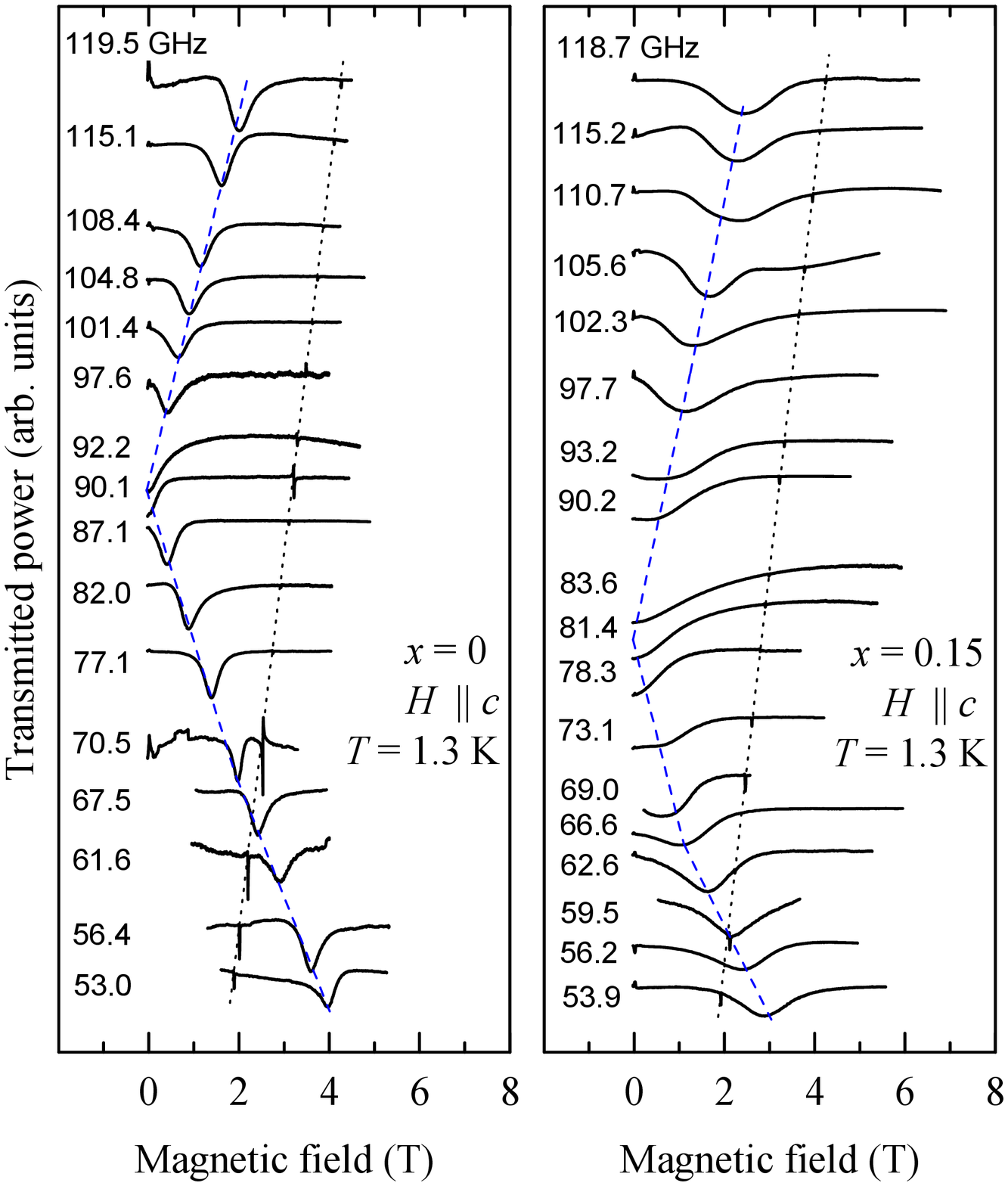}%%% FIG 9S %%%
 \caption{ Normalized ESR absorption curves for the \RKFM\ samples with $x=0$ and 0.15 for different frequencies.
 		Magnetic field is applied perpendicular to magnetic layers.
		Dashed lines are guide to the eye: blue lines are for antiferromagnetic resonance, black line is a DPPH marker.}
\label{fig:ESRlinesHperpplanex0&x0p15}
\end{figure}

\begin{figure}[tb]
\centering
\includegraphics[width=0.95\columnwidth]{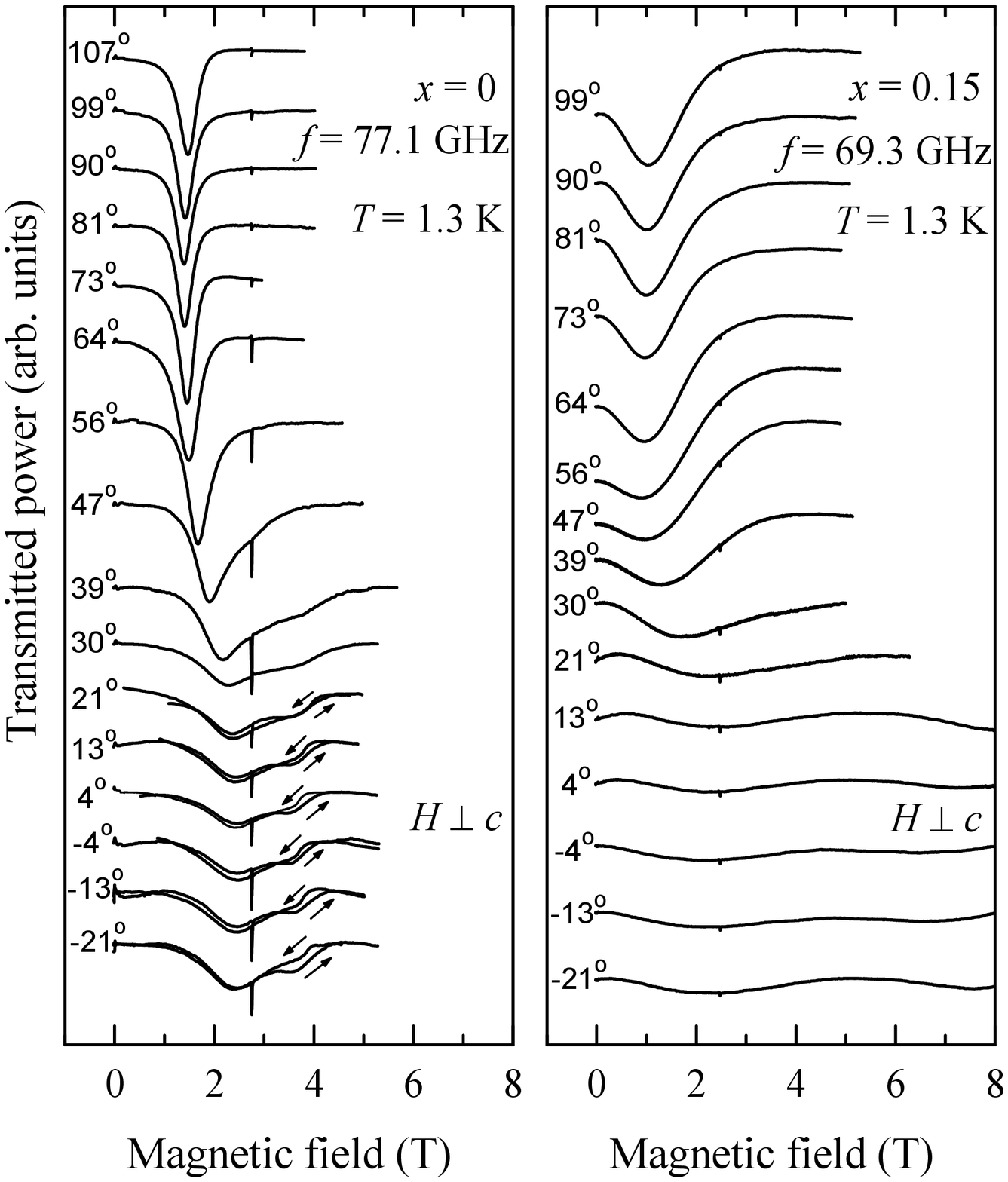}%%% FIG 10S %%%
\caption{Normalized ESR absorption curves for the \RKFM\ samples with $x=0$ and 0.15 for different orientations of the magnetic field on rotation from the in-plane to normal-to-plane direction.}
\label{fig:angular_dependence}
\end{figure}

For the magnetic field directed perpendicular to the planes of magnetic layers (i.e. ${\bf H} \parallel c$), we observe the ascending and descending branches both for the pure and the most highly doped $x=0.15$ samples, see Fig.~\ref{fig:ESRlinesHperpplanex0&x0p15}.
To confirm the phenomenon of the disappearance of the descending branch for ${\bf H} \perp c$, we studied the evolution of ESR records while gradually rotating the magnetic field from the ${\bf H} \parallel c$ orientation to the in-plane orientation at a frequency below the gap for the pure sample  and for the $x=0.15$ sample.
The results are presented in Fig.~\ref{fig:angular_dependence}.
One can see here, that for the doped sample the ESR line is smeared on rotation and is transformed into a curve which does not have a resonance shape and demonstrates only a weak variation with magnetic field.
At the same time, for the pure sample, the resonance absorption curve of the descending ESR mode survives the process of a rotation.
The increase of the linewidth of the pure sample, observed as a result of the rotation of the magnetic field to the in-plane direction may be naturally explained by the decrease of the absolute value of $df/dH$ from ${\bf H} \parallel c$ to ${\bf H} \perp c $, as observed in Ref.~[\onlinecite{PRB2007Smirnov}].

For the families of ESR curves presented in Figs.~\ref{fig:ESRlinesHinplanex0&x0p15}, \ref{fig:ESRlinesHperpplanex0&x0p15} and \ref{fig:angular_dependence} the gain is adjusted to keep the integrated intensity of the ESR signal at $T=10$~K equal for all frequencies.
This ensures that the records made at different frequencies and at different positions in the resonator are taken at the same sensitivity to microwave absorption in the sample.

The ESR absorption demonstrates a distinct difference in angular dependance of the lower branch for the pure and doped samples.
This shows the disappearance of the low-frequency absorption on rotation of the magnetic field in a doped sample, while for the pure sample this branch survives both for the in-plane and out-of plane fields.

\section{Theory}
%%%%%%%%%
We present here the calculation of the ESR spectrum for the conical/umbrella state which have been used to describe the experimental results for \RFM\ for magnetic fields applied along the $c$~axis.
We use the nearest-neighbor spin Hamiltonian without bond disorder:
\begin{equation}
\hat{\cal H} = J \sum_{\langle ij\rangle} {\bf S}_i\cdot {\bf S}_j + D \sum_i (S_i^z)^2 - g\mu_B\sum_i{\bf
H}\cdot{\bf S}_i \ .
\label{eq:H}
\end{equation}
For the easy-plane anisotropy and ${\bf H}\parallel \hat{\bf z}$ spins form the conical state described by the rotation angle $\theta_i = {\bf Q}\cdot{\bf R}_i$ in the $x$--$y$ plane and by the out of plane tilting angle $\alpha$.
The transformation between the laboratory frame $x_0,y_0,z_0$ and the rotating spin frame associated with the local sublattice orientation is given by
\begin{eqnarray}
S_i^{x_0} & = & -S_i^x\sin\theta_i + \cos\theta_i(S_i^z\cos\alpha-S_i^y\sin\alpha) \ , \nonumber \\
S_i^{y_0} & = & \phantom{-}S_i^x\cos\theta_i  + \sin\theta_i(S_i^z\cos\alpha-S_i^y\sin\alpha)\ , \\
S_i^{z_0} & = & \phantom{-}S_i^y\cos\alpha    + S_i^z\sin\alpha \ . \nonumber
\end{eqnarray}
The first two terms in the spin Hamiltonian (\ref{eq:H}) are expressed in the local frame as
\begin{eqnarray}
&& {\bf S}_i\cdot {\bf S}_j  =  -\frac{1}{2}S_i^zS_j^z(1-3\sin^2\alpha) -\frac{1}{2}S_i^xS_j^x + \nonumber \\
&& \mbox{} + S_i^yS_j^y(1-{\textstyle\frac{3}{2}}\sin^2\alpha) + (S_i^xS_j^y-S_i^yS_j^x)\sin\alpha\sin\theta_{ij} \\
&&\mbox{} + \frac{3}{4}(S_i^zS_j^y+S_i^yS_j^z)\sin 2\alpha+(S_i^zS_j^x-S_i^xS_j^z)\cos\alpha\sin\theta_{ij}\,,
\nonumber
\end{eqnarray}
where we have substituted $\cos\theta_{ij}= -1/2$, and
\begin{eqnarray}
(S_i^{z_0})^2 	& =	& (S_i^z)^2 \sin^2\alpha + (S_i^y)^2 \cos^2\alpha + \nonumber \\
			&	& \mbox{} + (S_i^zS_i^y+S_i^yS_i^z)\sin\alpha\cos\alpha \ .
\end{eqnarray}
Then, the classical energy is given by
\begin{equation}
E_{\rm cl}/N = -\frac{3}{2}JS^2(1-3\sin^2\alpha) + DS^2\sin^2\alpha - g\mu_BHS\sin\alpha \ .
\end{equation}
Minimization over the tilting angle yields $\sin\alpha =H/H_s$, where the saturation field $H_s$ is expressed as
\begin{equation}
g\mu_BH_s = 9JS + 2DS \ .
\end{equation}

We use the truncated Holstein-Primakoff transformation
\begin{eqnarray}
&& S_i^x = \sqrt{\frac{S}{2}}(a_i+a_i^\dagger) \ , \ \ S_i^y = -i\sqrt{\frac{S}{2}}(a_i-a_i^\dagger) \ , \nonumber \\
&& S_i^z = S - a_i^\dagger a_i
\end{eqnarray}
to obtain the harmonic spin-wave Hamiltonian. The second-order bosonic terms are
\begin{eqnarray}
&& \hat{\cal H}_2  = \sum_i \Bigl\{ \bigl[g\mu_B H\sin\alpha +DS(1-3\sin^2\alpha)\bigr] a_i^\dagger a_i \nonumber \\
&& \mbox{} -{\textstyle\frac{1}{2}}DS\cos^2\alpha(a_i^2+a_i^{\dagger 2}) \Bigr\} + JS \sum_{\langle ij\rangle} \Bigl\{ {\textstyle\frac{1}{2}} (1-3\sin^2\alpha) \nonumber \\
&& \times \bigl[a_i^\dagger a_i  + a_j^\dagger a_j +{\textstyle\frac{1}{2}}(a_i^\dagger a_j  + a_j^\dagger a_i)\bigr] - {\textstyle\frac{3}{4}}\cos^2\alpha (a_i a_j  + a_j^\dagger a_i^\dagger)  \nonumber \\
&& \phantom{J} - i\sin\alpha\sin\theta_{ij} (a_i^\dagger a_j  - a_j^\dagger a_i) \Bigr\} \ .
\end{eqnarray}
The Fourier transformed harmonic Hamiltonian takes the form
\begin{equation}
\hat{\cal H}_2 = \sum_{\bf k}\, (A_{\bf k}+C_{\bf k})\, a^\dagger_{\bf k} a_{\bf k} - \frac{1}{2} B_{\bf k}\, (a_{\bf k}a_{-\bf k} + a^\dagger_{\bf k} a^\dagger_{-\bf k}) \ ,
\end{equation}
where
\begin{eqnarray}
A_{\bf k} & = & 3JS + DS\cos^2\alpha + \frac{3}{2}\,JS\gamma_{\bf k} (1-3\sin^2\alpha) \ , \nonumber \\
B_{\bf k} & = & \frac{9}{2}\,JS \gamma_{\bf k}\cos^2\alpha  + DS\cos^2\alpha \ , \\
C_{\bf k} & = & 3JS\sin\alpha(\gamma_{\bf k+Q}- \gamma_{\bf k-Q})\phantom{\frac{3}{2}}\ , \nonumber
\end{eqnarray}
and $\gamma_{\bf k} = \frac{1}{3}(\cos k_x + 2 \cos(k_x/2)\cos(\sqrt{3}k_2/2)$. Note, that $A_{\bf k}$ and $C_{\bf k}$ are even and odd functions of the momentum $\bf k$, respectively.
Performing the standard Bogoliubov transformation we obtain the full magnon dispersion as
\begin{equation}
\epsilon_{\bf k}  = \sqrt{A_{\bf k}^2 - B_{\bf k}^2}  + C_{\bf k} \ .
\end{equation}

The ESR modes correspond to magnons  with momenta ${\bf k}=0$ and ${\pm\bf Q}$.
The former mode has zero energy $\epsilon_0\equiv 0$ related to the continuous rotational degeneracy about the field direction.
The other two modes give the ESR gaps
\begin{eqnarray}
\omega_1 & = & \frac{9}{2}JS \Bigl[\sqrt{\sin^2\alpha + (4D/9J)\cos^2\alpha} + \sin\alpha\Bigr] \ , \nonumber \\
\omega_2 & = & \frac{9}{2}JS \Bigl[\sqrt{\sin^2\alpha + (4D/9J)\cos^2\alpha} - \sin\alpha\Bigr]
\label{W12}
\end{eqnarray}
that have been used to fit the experimental data for ${\bf H} \parallel c$ both for pure and doped samples.

As noted in the main text, for pure samples the fitting parameters are  $J = 1.1J_0$ and $D = D_0$, and for $x=0.15$ sample $J = 0.9J_0$ ($D = D_0$), in a correspondence with a reduction of the exchange interaction
induced by K impurities.

%\bibliography{BibfileTLAFM}

\begin{thebibliography}{99}

\bibitem{Petrenko}
M. F. Collins and O. A. Petrenko, Can. J. Phys. {\bf 75}, 605 (1997).

\bibitem{Lewtas10}
H. J. Lewtas, A. T. Boothroyd, M. Rotter, D. Prabhakaran, H. M{\"u}ller, M. D. Le, B. Roessli, J. Gavilano, and
P. Bourges, Phys. Rev. B {\bf 82}, 184420 (2010).

\bibitem{Kenzelman07}
M. Kenzelmann, G. Lawes, A. B. Harris, G. Gasparovic, C. Broholm, A. P. Ramirez, G. A. Jorge, M. Jaime, S. Park,
Q. Huang, A. Ya. Shapiro, and L. A. Demianets, Phys. Rev. Lett. {\bf 98}, 267205 (2007).

\bibitem{Lee84}
D. H. Lee, J. D. Joannopoulos, J. W. Negele, and D. P. Landau, Phys. Rev. Lett. {\bf 52}, 433 (1984).

\bibitem{Kawamura}
H. Kawamura and  S. Miyashita, J. Phys. Soc. Jpn. {\bf 54}, 4530 (1985).

\bibitem{Chubukov91}
A. V. Chubukov and D. I. Golosov, J. Phys.: Condens. Mat. {\bf 3}, 69 (1991).

\bibitem{Inami96}
T. Inami, Y. Ajiro, and T. Goto, J. Phys. Soc. Jpn. {\bf 65}, 2374 (1996).

\bibitem{SvistovPRB2003}
L. E. Svistov, A. I. Smirnov, L. A. Prozorova, O. A.  Petrenko, L. N. Demianets,  and A. Ya. Shapiro, Phys. Rev.
B {\bf 67}, 094434 (2003).

\bibitem{Kitazawa99}
H. Kitazawa, H. Suzuki, H. Abe, J. Tang, and G. Kido, Physica B {\bf 259-261}, 890 (1999).

\bibitem{Ishii11}
R. Ishii, S. Tanaka, K. Onuma, Y. Nambu, M. Tokunaga, T. Sakakibara, N. Kawashima, Y. Maeno, C. Broholm, D. P.
Gautreaux, J. Y. Chan, and S. Nakatsuji, EPL {\bf 94}, 17001 (2011).

\bibitem{ShirataPRL}
Yu. Shirata, H. Tanaka, A.  Matsuo,  and K. Kindo, Phys. Rev. Lett. {\bf 108}, 057205 (2012).

\bibitem{Ueda05}
H. Ueda, H. A. Katori, H. Mitamura, T. Goto, and H. Takagi, Phys. Rev. Lett. {\bf 94}, 047202 (2005).

\bibitem{Penc04}
K. Penc, N. Shannon, and H. Shiba, Phys. Rev. Lett. {\bf 93}, 197203 (2004).

\bibitem{Ueda06}
H. Ueda,  H. Mitamura, T. Goto, and Y. Ueda, Phys. Rev. B {\bf 73}, 094415 (2006).

\bibitem{Maryasin13}
V. S. Maryasin and M. E. Zhitomirsky, Phys. Rev. Lett. {\bf 111}, 247201 (2013).

\bibitem{Henley89}
C. L. Henley, Phys. Rev. Lett. {\bf 62}, 2056 (1989).

\bibitem{Long89}
M. W. Long, J. Phys.: Condens. Mat. {\bf 1}, 2857 (1989).

\bibitem{Fyodorov91}
Y. V. Fyodorov and E. F. Shender, J. Phys. Condens. Matter {\bf 3}, 9123 (1991).

\bibitem{Maryasin15}
V. S. Maryasin and M. E. Zhitomirsky, J. Phys.: Confer. Ser. {\bf 592}, 012112 (2015).

\bibitem{Slon91}
J. C. Slonczewski, Phys. Rev. Lett. {\bf 67}, 3172 (1991).

\bibitem{Demokritov98}
S O. Demokritov, J. Phys. D {\bf 31}, 925 (1998).

\bibitem{Schmidt1999}
C. M. Schmidt,   D. E. B\"urgler,   D. M. Schaller,  F.Meisinger,  and H.-J. G\"untherodt, Phys. Rev. B {\bf 60},
4158 (1999).

\bibitem{Svistov2006}
L. E. Svistov, A. I. Smirnov, L. A. Prozorova, O. A.  Petrenko, A. Micheler, N. B\"uttgen, A. Ya. Shapiro, and L.
N. Demianets, Phys. Rev. B {\bf 74}, 024412 (2006).

\bibitem{Smirnov2007}
A. I. Smirnov, H. Yashiro, S. Kimura, M. Hagiwara, Y. Narumi, K. Kindo, A. Kikkawa, K. Katsumata,  A. Ya.
Shapiro, and L. N. Demianets, Phys. Rev. B {\bf 75}, 134412 (2007).

\bibitem{KFeMoOPRL}
A. I. Smirnov, L. E. Svistov, L. A. Prozorova, A. Zheludev, M. D. Lumsden, E. Ressouche,  O. A.  Petrenko, K. Nishikawa, S.
Kimura, M. Hagiwara, K. Kindo, A. Ya. Shapiro, and L. N. Demianets, Phys. Rev. Lett. {\bf 102}, 037202 (2009).

\bibitem{Supplemental_material}
See Supplemental Material for further experimental and theoretical
details.

\bibitem{White13}
J. S. White, Ch. Niedermayer, G. Gasparovic, C. Broholm, J. M. S. Park,  A. Ya. Shapiro, L. N.
Demianets, and M. Kenzelmann, Phys. Rev. B {\bf 88}, 060409 (2013).

\bibitem{MZH15}
M. E. Zhitomirsky, J. Phys.: Confer. Ser. {\bf 592}, 012110 (2015).

\bibitem{Shender82}
E. F. Shender, Zh. Eksp. Teor. Fiz. {\bf 83}, 326 (1982) [Sov. Phys. JETP {\bf 56}, 178 (1982)].

\bibitem{Moessner01}
R. Moessner, Can. J. Phys. {\bf 79}, 1283, (2001).

\end{thebibliography}

\begin{thebibliography}{10}
\bibitem{PRB2003Svistov} L. E. Svistov, A. I. Smirnov, L. A. Prozorova, O. A. Petrenko, L. N. Demianets, and A. Ya. Shapiro, 
 Phys. Rev. B {\bf 67}, 094434 (2003).

\bibitem{PRB2007Smirnov} A. I. Smirnov, H. Yashiro, S. Kimura, M. Hagiwara, Y. Narumi, K. Kindo, A. Kikkawa, K. Katsumata, A. Ya.
Shapiro, and L. N. Demianets, Phys. Rev. B {\bf 75}, 134412 (2007).
\end{thebibliography}

\end{document}